\documentclass{ws-ijmpa}

\begin{document}

\markboth{N.Sawado}
{Hopf solitons solutions from low energy effective action of SU(2) 
Yang-Mills theory}

%
%

\title{HOPF SOLITON SOLUTIONS FROM LOW ENERGY EFFECTIVE ACTION OF SU(2) 
YANG-MILLS THEORY}

\author{N.Sawado}
\address{Department of Physics, Faculty of Science and Technology, 
Tokyo University of Science, Noda, Chiba 278-8510, Japan 
\\
sawado@ph.noda.tus.ac.jp}

\author{N.Shiiki}
\address{Department of Physics, Faculty of Science and Technology,
Tokyo University of Science, Noda, Chiba 278-8510, Japan
\\
norikoshiiki@mail.goo.ne.jp}

\author{S.Tanaka}
\address{Department of Physics, Faculty of Science and Technology,
Tokyo University of Science, Noda, Chiba 278-8510, Japan}

\maketitle

\begin{history}
\end{history}

\begin{abstract}
The Skyrme-Faddeev-Niemi (SFN) model which is an O(3) $\sigma$ model in three 
dimensional space up to fourth-order in the first derivative is regarded as a 
low-energy effective theory of SU(2) Yang-Mills theory.
One can show from the Wilsonian renormalization group argument that 
the effective action of Yang-Mills theory recovers the SFN in the infrared 
region. However, the theory contains another fourth-order term 
which destabilizes the soliton solution. 
In this paper we derive an extended action including second derivative 
terms and obtain soliton solutions numerically. A new topological lower bound 
formula is infered for the extended action. 
\keywords{Topological solitons; Yang-Mills theory; Second derivative field theory.}
\end{abstract}

\ccode{PACS numbers: 11.10.Lm, 11.27.+d, 12.38.Aw, 12.38.Lg, 12.39.Dc}

\section{Introduction}	

The Skyrme-Faddeev-Niemi (SFN) model which is an O(3) $\sigma$ model 
in three dimensional space up to fourth-order in the first derivative 
possesses topological soliton solutions with torus or knot-like structure. 
The model was initiated in 70's~\cite{faddeev75} and its interest has 
been extensively growing. The numerical simulations were performed 
\cite{faddeev97,gladikowski97,sutcliffe98,hietarinta99,hietarinta00}, 
the integrability was shown\cite{aratyn99}, and the application to 
the condensed matter physics~\cite{babaev02} and the Weinberg-Salam 
model~\cite{fayzullaev} were also considered. 
The recent research especially focuses on the consistency between 
the SFN and fundamental theories such as QCD~\cite{faddeev99,langmann99,shabanov99,cho02}. 
In those references, it is claimed that the SFN action should be derived    
from the SU(2) Yang-Mills (YM) action at low energies.
One can also show from the Wilsonian renormalization group argument that 
the effective action of Yang-Mills theory recovers the SFN in the infrared 
region~\cite{gies01}. 
However, the derivative expansion for slowly varying 
fields $\bm{n}$ up to quartic order produces an additional fourth-order term 
in the SFN model, resulting in instability of the soliton solution. 

Similar situations can be seen also in various topological soliton 
models. In the Skyrme model, the chirally invariant lagrangian 
with quarks exhibits fourth order terms after the derivative expansion 
and they destabilize the soliton solution~\cite{dhar85,aitchison85}.
To recover the stability of the skyrmion, the author of Ref.\refcite{marleau01} 
introduced a large number of higher order terms in the first derivative 
whose coefficients were determined from the coefficients of the 
Skyrme model by using the recursion relations. 
Alternatively, in Ref.~\refcite{gies01} Gies pointed out the possibility that 
the second derivative order term can work as a stabilizer for the soliton. 
Similar form of the action was also proposed by Forkel 
with somewhat different expansion scheme\cite{forkel05}. The author also found 
out the saddle point soliton solutions with hedgehog type. This result is quite 
encouraging to examine the consistency between the topological soliton 
physics and the Yang-Mills theory. 

In this paper, we compute the extended Hopf soliton solutions from 
the action proposed by Gies. In section \ref{sec:level2}, 
we give an introduction to the Skyrme-Faddeev-Niemi 
model with its topological property. 
In section \ref{sec:level3}, we show how to derive the SFN model action 
from the SU(2) Yang-Mills theory. In section \ref{sec:level4}, 
soliton solutions of this truncated YM action are studied. 
For this purpose, we introduce a second derivative 
term which can be derived in a perturbative manner.
The naive extremization scheme, however, produces the fourth 
order differential equation and the model has no stable soliton solution. 
Failure of finding the soliton is caused by the 
basic feature of the second derivative field theory.
In section \ref{sec:level5}, the higher derivative theory and 
Ostrogradski's formulation are reviewed. We show the absence 
of the energy bound in the second derivative theory using an example in 
quantum mechanics and introduce the perturbative treatment for the 
second derivative theory. 
In section \ref{sec:level6}, we present our numerical results. 
The possibility of finding new topological bound for this extended, 
perturbative soliton solutions is also discussed. 
In section \ref{sec:level7} are concluding remarks. 

\section{\label{sec:level2}Skyrme-Faddeev-Niemi model}

The Faddeev-Niemi conjecture for the low-energy model of SU(2) 
Yang-Mills theory is expressed by the following effective action:
\begin{eqnarray}
S_{\rm SFN}=\Lambda \int d^4x\Bigl[\frac{1}{2}(\partial_\mu \bm{n})^2
+\frac{g_1}{8}(\partial_\mu \bm{n}\times \partial_\nu \bm{n})^2 \Bigr]\label{fsn_ac}
\end{eqnarray}
where $\bm{n}(\bm{x})$ is a three component vector field normalized as 
$\bm{n}\cdot\bm{n}=1$. The mass scale $\Lambda$ is a free parameter 
and in this paper we set $\Lambda=1$. It has been shown that 
stable soliton solutions exist when $g_1 > 0$.

The static field $\bm{n}(\bm{x})$ maps $\bm{n}:R^3\mapsto S^2$ and 
the configurations are classified by the topological maps characterized 
by a topological invariant $H$ called Hopf charge 
\begin{eqnarray}
H=\frac{1}{32\pi^2}\int A \wedge F,~~F=dA
\label{hopf}
\end{eqnarray}
where $F$ is the field strength and expressed in terms of 
$\bm{n}(\bm{x})$ as $F=(\bm{n}\cdot d\bm{n}\wedge d\bm{n})$.

The static energy $E_{\rm stt}$ from the action (\ref{fsn_ac}) has 
a topological lower bound~\cite{ward98}, 
\begin{eqnarray}
E_{\rm stt}\ge K H^{3/4}
\label{lowerbound}
\end{eqnarray} 
where $K=16\pi^2\sqrt{g_1}$.

Performing numerical simulations, one can find that the static configurations 
for $H=1,2$ have axial symmetry~\cite{sutcliffe98}. 
Thus ``the toroidal ansatz'' is suitable to impose on these 
configurations\cite{gladikowski97}. The ansatz is given by 
\begin{eqnarray}
&&n_1=\sqrt{1-w^2(\eta,\beta)}\cos(N\alpha+v(\eta,\beta))\,, \nonumber \\
&&n_2=\sqrt{1-w^2(\eta,\beta)}\sin(N\alpha+v(\eta,\beta))\,, 
\label{toroidal} \\
&&n_3=w(\eta,\beta)\,, \nonumber 
\end{eqnarray}
where $(\eta,\beta,\alpha)$ is toroidal coordinates which are related to 
the Gaussian coordinates in $R^3$ as follows:
\begin{eqnarray}
x=\frac{a\sinh\eta\cos\alpha}{\tau},y=\frac{a\sinh\eta\sin\alpha}{\tau},
z=\frac{a\sin\beta}{\tau}
\end{eqnarray}
with $\tau=\cosh\eta-\cos\beta$.
The function $w(\eta,\beta)$ is subject to the boundary conditions $w(0,\beta)=1,w(\infty,\beta)=-1$
and is periodic in $\beta$. $v(\eta,\beta)$ is set to be $v(\eta,\beta)=M\beta+v_0(\eta,\beta)$ and 
$v_0(,\beta)$ is considered as a constant map. 
Equation (\ref{hopf}) then gives $H=NM$.

To obtain soliton solutions with higher derivative terms, we propose 
a simpler ansatz than 
(\ref{toroidal}), which 
is defined by
\begin{eqnarray}
&&n_1=\sqrt{1-w^2(\eta)}\cos(N\alpha+M\beta)\,,\nonumber \\
&&n_2=\sqrt{1-w^2(\eta)}\sin(N\alpha+M\beta)\,, 
\label{afz} \\
&&n_3=w(\eta)\,, \nonumber 
\end{eqnarray}
where $w(\eta)$ satisfies the boundary conditions $w(0)=1,w(\infty)=-1$. 
By using (\ref{afz}), the static energy is written in terms of the function 
$w(\eta)$ as
\begin{eqnarray}
&&E_{\rm stt}=2\pi^2a \int d\eta
\biggl[
\frac{(w')^2}{1-w^2}+(1-w^2) U_{M,N}(\eta)
+\frac{g_1}{2a^2}\sinh\eta\cosh\eta (w')^2
U_{M,N}(\eta)\biggr]\,,\nonumber \\
&&\hspace{1cm}w'\equiv \frac{dw}{d\eta},~~
U_{M,N}(\eta)\equiv \Bigl(M^2+\frac{N^2}{\sinh^2\eta}\Bigr)\,. \nonumber
\end{eqnarray}
The Euler-lagrange equation of motion is then derived as
\begin{eqnarray}
&&\frac{w''}{1-w^2}+\frac{ww'^2}{(1-w^2)^2}+U_{M,N}(\eta)w
+\frac{g_1}{2a^2}\Bigl(-2N^2\coth^2\eta w'\nonumber \\
&&+(\cosh^2\eta+\sinh^2\eta)
U_{M,N}(\eta)w'+\sinh\eta\cosh\eta U_{M,N}(\eta)w''\Bigr)=0\,.
\label{fsn_eq}
\end{eqnarray}
The variation with respect to $a$ produces the equation for $a$.
Soliton solutions are obtained by solving the equations for $a$ as 
well as for $w$.

\begin{figure}
\includegraphics[height=9cm, width=12cm]{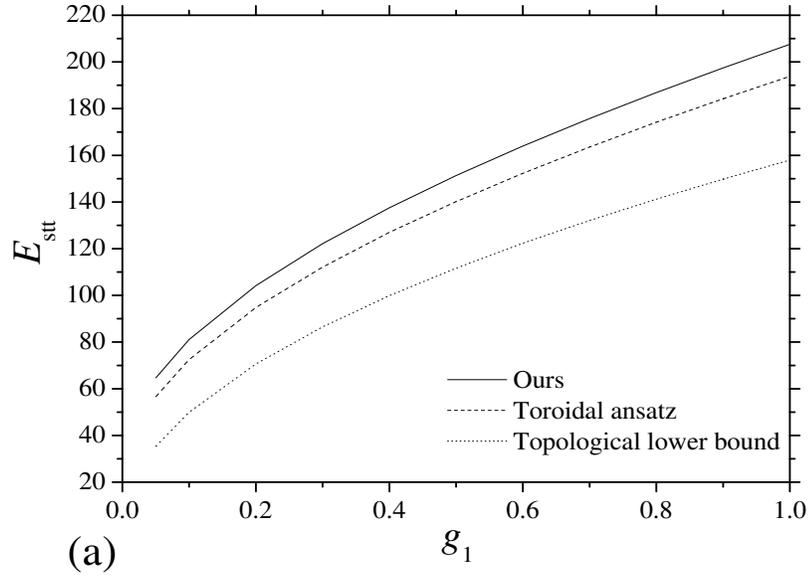}
\includegraphics[height=9cm, width=12cm]{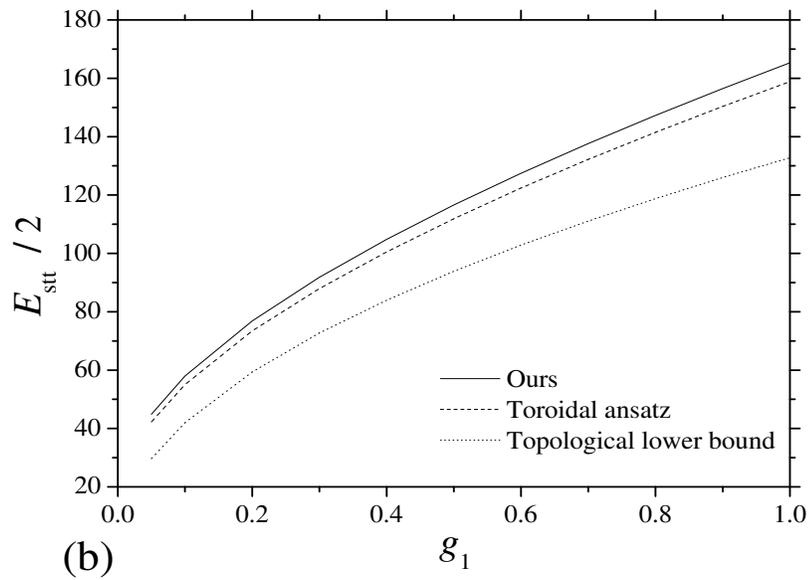}
\caption{\label{fig:Fig1}The total energy of the solitons with (a) $H=1$, (b) $H=2$, 
as a function of the coupling constant $g_1$: the simple ansatz (\ref{afz}), 
the toroidal ansatz (\ref{toroidal}) by Gladikowski-Hellmund, and the expected 
topological lower bound (\ref{lowerbound}).}
\end{figure}

We obtained soliton solutions numerically for both ansatz (\ref{toroidal})
and (\ref{afz}). The total energies together with the topological lower 
bound (\ref{lowerbound}) are shown as a function of coupling 
constant $g_1$ in Fig.\ref{fig:Fig1}. We found that this simple 
ansatz produces at most 10\% errors and it does not affect to 
the property of the soliton solution.

\section{\label{sec:level3}Cho-Faddeev-Niemi-Shabanov decomposition and 
the effective action in the Yang-Mills theory\protect\\}

In this section, we briefly review how to derive the 
SFN effective action from the action of SU(2) Yang-Mills theory in the 
infrared limit~\cite{shabanov99,gies01}. 
For the gauge fields $\bm{A}_\mu$, the Cho-Faddeev-Niemi-Shabanov 
decomposition is applied~\cite{faddeev99,langmann99,shabanov99,cho02}

\begin{eqnarray}
\bm{A}_\mu=\bm{n}C_\mu+(\partial_\mu\bm{n})\times\bm{n}+\bm{W}_\mu\,.
\label{cfns}
\end{eqnarray}
The first two terms are the ``electric'' and ``magnetic'' Abelian connection,
and $\bm{W}_\mu$ are chosen so as to be orthogonal to $\bm{n}$, {\it i.e.} 
$\bm{W}_\mu\cdot\bm{n}=0$.
Obviously, the degrees of freedom on the left- and right-hand side 
of Eq.(\ref{cfns}) do not match. While the LHS describes 
$3_{\rm color}\times 4_{\rm Lorentz}=12$, the RHS consists of 
$(C_\mu:)4_{\rm Lorentz}+(\bm{n}:)2_{\rm color}+(\bm{W}_\mu:)3_{\rm color}
\times4_{\rm Lorentz}-4_{\bm{n}\cdot\bm{W}_\mu=0}=14$ degrees freedom. 
Shabanov introduced in his paper \cite{shabanov99} the following constraint 
\begin{eqnarray}
\bm{\chi}(\bm{n},C_\mu,\bm{W}_\mu)=0,~{\rm with}~~\bm{\chi}\cdot\bm{n}=0\,.
\end{eqnarray}
The generating functional of YM theory can be written as
\begin{eqnarray}
{\cal Z}=\int {\cal D}\bm{n}{\cal D}C{\cal D}\bm{W}\delta(\bm{\chi})
\Delta_{\rm FP}\Delta_{\rm S}e^{-S_{\rm YM}-S_{\rm gf}}\,.
\label{vf0}
\end{eqnarray}
$\Delta_{\rm FP}$ and $S_{\rm gf}$ are the Faddeev-Popov determinant and 
the gauge fixing action respectively, and Shabanov introduced another 
determinant $\Delta_{\rm S}$ corresponding to the condition $\bm{\chi}=0$.
YM and the gauge fixing action is given by 
\begin{eqnarray}
&&S_{\rm YM}+S_{\rm  gf}=\int d^4x\Bigl[\frac{1}{4g^2}\bm{F}_{\mu\nu}\cdot\bm{F}_{\mu\nu}
+\frac{1}{2\alpha_{\rm g} g^2}(\partial_\mu \bm{A}_\mu)^2\Bigr] \,. \nonumber 
\end{eqnarray}
Inserting Eq.(\ref{cfns}) into the action, one obtains the vacuum functional
\begin{eqnarray}
&&{\cal Z}=\int {\cal D}\bm{n}e^{-{\cal S}_{\rm eff}(\bm{n})}
=\int {\cal D}\bm{n} e^{-{\cal S}_{\rm cl}(\bm{n})}\int {\cal D}\bar{C}{\cal D}\bar{\bm{W}}_\mu
\Delta_{\rm FP}\Delta_{\rm S}\delta({\bm \chi}) \nonumber \\
&&~~~\times e^{-(1/2g^2)\int[\bar{C}_\mu M^C_{\mu\nu}\bar{C}_\nu+\bar{\bm{W}}_\mu\cdot\bar{M}^{\bm{W}}_{\mu\nu}
\bar{\bm{W}}_\nu-K_\mu^C(M_{\mu\nu}^C)^{-1}K_\nu^C
+\bar{\bm{K}}_\mu^{\bm{W}}\cdot (\bar{M}_{\mu\nu}^{\bm{W}})^{-1} 
\bar{\bm{K}}_\nu^{\bm{W}}]}
\end{eqnarray}
with 
\begin{eqnarray}
&&\bar{M}^{\bm{W}}_{\mu\nu}:=M^{\bm{W}}_{\mu\nu}+\bm{Q}_{\mu s}(M^C_{s\lambda})^{-1}\bm{Q}_{\lambda \nu}\,,~~
\bar{\bm{W}}_\mu=\bm{W}_\mu-(\bar{M}_{\mu\nu}^{\bm{W}})^{-1}K_\nu^{\bm{W}}\,, 
\nonumber \\ 
&&\hspace{1cm}\bar{C}_\mu=C_\mu+(M^C_{\mu\lambda})^{-1}
\bm{Q}_{\lambda \nu}\cdot\bm{W}_\nu+(M_{\mu\nu}^C)^{-1}K_\nu^C\,,
\end{eqnarray}
and
\begin{eqnarray}
&&M^C_{\mu\nu}=-\partial^2\delta_{\mu\nu}+\partial_\mu\bm{n}\cdot\partial_\nu
\bm{n}\,,~~
M^{\bm{W}}_{\mu\nu}=-\partial^2\delta_{\mu\nu}-\partial_\mu\bm{n}\otimes\partial_\nu\bm{n}
+\partial_\nu\bm{n}\otimes\partial_\mu\bm{n}\,, \nonumber \\
&&\bm{Q}^C_{\mu\nu}=\partial_\mu\bm{n}\partial_\nu
+\partial_\nu\bm{n}\partial_\mu+\partial_\mu\partial_\nu\bm{n}\,,~~
K^C_{\mu\nu}=\partial_\nu(\bm{n}\cdot\partial_\nu\bm{n}\times\partial_\mu\bm{n})
+\partial_\mu\bm{n}\cdot\partial^2\bm{n}\times\bm{n}\,,
\nonumber \\
&&\hspace{2.5cm}\bm{K}^{\bm W}_{\mu}=\partial_\mu(\bm{n}\times\partial^2\bm{n})\,,
~~{\rm (in~gauge~\alpha_g=1})\,.
\end{eqnarray}
The classical action for $\bm{n}$ including the gauge fixing term is given by
\begin{eqnarray}
{\cal S}_{\rm cl}=\int d^4x\Bigl[\frac{1}{4g^2}(\partial_\mu\bm{n}\times \partial_\nu\bm{n})^2
+\frac{1}{2 \alpha_{\rm g} g^2}(\partial^2\bm{n}\times\bm{n})^2\Bigr]\,.
\end{eqnarray}
The $\delta$ functional is expressed by its Fourier transform
\begin{eqnarray}
\delta(\bm{\chi})=\int {\cal D}\bm{\phi}e^{-i\int (\bm{\phi}\cdot \partial
\bm{W}_\mu
+\bm{\phi}\cdot C_\mu\bm{n}\times \bm{W}_\mu+(\bm{\phi}\cdot\bm{n})(\partial_\mu\bm{n}\cdot\bm{W}_\mu))}\,.
\end{eqnarray} 
Integrating over ``fast'' variables $C,\bm{W},\bm{\phi}$, we finally obtain 
\begin{eqnarray}
&&e^{-S_{\rm eff}}=e^{-S_{\rm cl}}\Delta_{\rm FP}\Delta_{\rm S}
(\det M^C)^{-1/2} (\det \bar{M}^{\bm{W}})^{-1/2}
(\det -Q^{\bm{\phi}}_\mu (\bar{M}^{\bm{W}})^{-1}_{\mu\nu}
Q^{\bm{\phi}}_\nu)^{-1/2} \nonumber \\
\label{determ} 
\end{eqnarray}
where $Q_\mu^\phi:=i(-\partial_\mu+\partial_\mu\bm{n}\otimes\bm{n})$.
Here several nonlocal terms and the higher derivative components have been 
neglected. 

Performing the derivative expansion for the resultant determinants 
with $\partial\bm{n}$, one obtain the effective action for ``slow''variable $\bm{n}$\cite{gies01}
\begin{eqnarray}
&&S_{\rm eff}=\int d^4x\Bigl[\frac{1}{2}(\partial_\mu \bm{n})^2
+\frac{g_1}{8}(\partial_\mu \bm{n}\times \partial_\nu \bm{n})^2
+\frac{g_2}{8}(\partial_\mu \bm{n})^4\Bigr]\,.
\label{fsn2}
\end{eqnarray}
For $g_1>0$ and $g_2=0$, the action is identical to the SFN effective 
action~(\ref{fsn_ac}).

In order for soliton solutions to exist, $g_2$ must be 
positive~\cite{gladikowski97}. However, $g_2$ is found to be 
negative according to the above analysis.
Therefore we consider higher-derivative terms and investigate 
if the model with the higher-derivatives possess soliton solutions. 

\section{\label{sec:level4}Search for the soliton solutions (1)
\protect\\}
The static energy is derived from Eq.(\ref{fsn2}) as 
\begin{eqnarray}
E_{\rm stt}&=& \int d^3x\Bigl[\frac{1}{2}(\partial_i \bm{n})^2
+\frac{g_1}{8}(\partial_i \bm{n}\times \partial_j \bm{n})^2
+\frac{g_2}{8}(\partial_i \bm{n})^4 \Bigr] \nonumber \\
&:=&E_2(\bm{n})+E_4^{(1)}(\bm{n})+E_4^{(2)}(\bm{n})\,.
\end{eqnarray}
A spatial scaling behaviour of the static energy, so called Derrick's scaling 
argument, can be applied to examine the stability of the 
soliton~\cite{sutcliffe05}. 
Considering the map 
$\bm{x}\mapsto \bm{x}'=\mu\bm{x}~(\mu>0)$, with 
$\bm{n}^{(\mu)}\equiv \bm{n}(\mu\bm{x})$, the static energy scales as 
\begin{eqnarray}
e(\mu)&=&E_{\rm stt}(\bm{n}^{(\mu)}) \nonumber \\
&=&E_2(\bm{n}^{(\mu)})+E_4^{(1)}(\bm{n}^{(\mu)})+E_4^{(2)}(\bm{n}^{(\mu)}) 
\nonumber \\
&=&\frac{1}{\mu}E_2(\bm{n})+\mu(E_4^{(1)}(\bm{n})+E_4^{(2)}(\bm{n}))\,.
\label{derrick}
\end{eqnarray}
Derrick's theorem states that if the function $e(\mu)$ has no stationary 
point, the theory has no static solutions of the field equation with finite 
density other than the vacuum. 
Conversely, if $e(\mu)$ has stationary point, the  possibility of 
having finite energy soliton solutions is not excluded. 
Eq.(\ref{derrick}) is stationary at $\mu=\sqrt{E_2/(E_4^{(1)}+E_4^{(2)})}$. 
Then, the following inequality
\begin{eqnarray}
&&g_1(\partial_i\bm{n}\times\partial_j\bm{n})^2+g_2(\partial_i\bm{n})^2(\partial_j\bm{n})^2 \nonumber \\
&&=g_1(\partial_i\bm{n})^2(\partial_j\bm{n})^2-g_1(\partial_i\bm{n}\cdot\partial_j\bm{n})^2
+g_2(\partial_i\bm{n})^2(\partial_j\bm{n})^2 \nonumber \\
&&\geqq g_2(\partial_i\bm{n}\cdot\partial_j\bm{n})^2
~~~~(\because (\partial_i\bm{n})^2(\partial_j\bm{n})^2\geqq 
(\partial_i\bm{n}\cdot\partial_j\bm{n})^2) \nonumber
\end{eqnarray}
ensures the possibility of existence of the stable soliton solutions for 
$g_2\geqq 0$. 
As mentioned in the section \ref{sec:level3}, $g_2$ should be negative at 
least within our derivative expansion analysis of YM theory. 

A promising idea to tackle the problem was suggested by Gies~\cite{gies01}.
The author considered the following type of effective action, accompanying 
an second derivative term 
\begin{eqnarray}
&&S_{\rm eff}=\int d^4x\Bigl[\frac{1}{2}(\partial_\mu \bm{n})^2
+\frac{g_1}{8}(\partial_\mu \bm{n}\times \partial_\nu \bm{n})^2 
-\frac{g_2}{8}(\partial_\mu \bm{n})^4
+\frac{g_2}{8}(\partial^2 \bm{n}\cdot\partial^2 \bm{n})
 \Bigr]\,.
 \label{fsn_ac2}
 \end{eqnarray}
Here we choose positive value of $g_2$ and assign the explicit negative sign 
to the third term. In principle, it is possible to estimate the second 
derivative term by the derivative expansion without neglecting 
throughout the calculation.
The calculation is, however, very laborious and hence we will show in detail 
somewhere else. 

The static energy of Eq.(\ref{fsn_ac2}) with the ansatz (\ref{afz}) is 
written as 
\begin{eqnarray}
&&E_{\rm stt}=2\pi^2a \int d\eta
\Biggl[
\frac{(w')^2}{1-w^2}+(1-w^2) U_{M,N}(\eta)
+\frac{g_1}{2a^2}\sinh\eta\cosh\eta (w')^2
U_{M,N}(\eta)\nonumber \\
&&+\frac{g_2}{4a^2}\biggl[-\sinh\eta\cosh\eta\Bigl[\frac{(w')^2}{1-w^2}
+(1-w^2)U_{M,N}(\eta)\Bigr]^2 \nonumber \\
&&+\Bigl(\coth\eta+\sinh^2\eta-\sinh\eta\cosh\eta\Bigr)
\frac{(w')^2}{1-w^2} \nonumber \\
&&+(\sinh\eta\cosh\eta-\sinh^2\eta)(1-w^2)M^2 
+2\Bigl\{ \frac{w(w')^3}{(1-w^2)^2}
+\frac{w' w''}{1-w^2}
+w w'U_{M,N}(\eta) \Bigr\} \nonumber \\
&&+\sinh\eta\cosh\eta\Bigl\{ \frac{1}{1-w^2}\Bigl[\frac{(w')^2}{1-w^2}
+w w''
+(1-w^2)U_{M,N}(\eta)\Bigr]^2 
+(w'')^2\Bigr\}\biggr]\Biggl],
\label{stene2}
\end{eqnarray} 
where $w''\equiv \frac{d^2w}{d\eta^2}$.
The Euler-Lagrange equation of motion is derived by 
\begin{eqnarray}
-\frac{d^2}{d\eta^2}\Bigl(\frac{\partial E_{\rm stt}}{\partial w''}\Bigr)
+\frac{d}{d\eta}\Bigl(\frac{\partial E_{\rm stt}}{\partial w'}\Bigr)
-\frac{\partial E_{\rm stt}}{\partial w}=0 \,,
\end{eqnarray}
which is too complicated to write explicitly. Thus we adopt the following 
notation  
\begin{eqnarray}
f_0(w,w',w'')+g_1f_1(w,w',w'')+g_2f_2(w,w',w'',w^{(3)},w^{(4)})=0\,.
\label{fsn_eq2}
\end{eqnarray}
Here $w^{(3)},w^{(4)}$ represent the third and the fourth derivative with 
respect to $\eta$. The first two terms of Eq.(\ref{fsn_eq2}) are 
identical to those in Eq.(\ref{fsn_eq}).

In addition to the boundary conditions at the origin and the infinity 
$\omega(0)=1,\omega(\infty)=-1$, 
the regularity condition specifies $\omega'(0)=\omega''(0)=0$~(see 
Eq.(\ref{stene2})).

Unfortunately, under these boundary conditions we could not find 
soliton solutions in Eq.(\ref{fsn_eq2}) for any value of $g_2$. 

Recently, Forkel studied the soft mode action of the Yang-Mills theory and 
investigated the solitonic, saddle point solution with hegdehog symmetry
\cite{forkel05}. His work is quite encouraging to find the class of 
solutions in the infrared part of the Yang-Mills action. 
But, the solutions the author obtained are solitons with non-zero topological 
charge and, in the case of the Hopf solitons which possess zero topological charge, 
the situation is a little more complicated. 

From the identity 
\begin{eqnarray}
\int d^4x[(\partial^2 \bm{n}\cdot\partial^2 \bm{n})-(\partial_\mu \bm{n})^4]
=\int d^4x(\partial^2 \bm{n}\times\bm{n})^2\,,
\end{eqnarray} 
one easily finds that the static energy obtained from the last two terms 
in Eq.(\ref{fsn_ac2}) 
\begin{eqnarray}
\tilde{E}^{(2)}_4=\int d^3x(\partial^2 \bm{n}\times\bm{n})^2
\label{energy2}
\end{eqnarray}
gives the positive contribution. The total static energy is stationary 
at $\mu=\sqrt{E_2/(E_4^{(1)}+\tilde{E}_4^{(2)})}$ and hence the possibility 
of existence of soliton solutions is not excluded. 
And also, the positivity of Eq.(\ref{energy2}) does not spoil the lower 
bound (\ref{lowerbound}) of original SFN and the possibility still remains.  

Therefore, we suspect that the failure of finding the stable soliton 
is caused by the fact that {\it higher derivative theory has no lower bound}. 
We shall investigate the lower bound in the higher derivative theory 
in detail in the next section. 

\section{\label{sec:level5}Higher derivative theory\protect\\}
In this section, we address the basic problems in the higher derivative 
theory~\cite{pais,smilga,eliezer89,jaen,simon} which essentially falls 
into two categories. The first problem concerns the increase in the 
number of degrees of freedom. For example, if the theory contains 
second derivative terms, the equation of motion becomes up to the order 
in the fourth derivative. Thus, four parameters are required for 
the initial conditions. If one considers more higher order terms, 
the situation gets worse. However, this is not serious problem for 
our study because our concern is the existence of static soliton solutions. 
The second problem is that the actions of the theory are not bounded from 
below. This feature makes the higher derivative theories unstable.

The lagrangian and the hamiltonian formalism with higher derivative 
was firstly developed by Ostrogradski~\cite{ostrogradski}. We consider 
the lagrangian containing up to $n$th order derivatives 
\begin{eqnarray}
S=\int dt {\cal L}(q,\dot{q},\cdots,q^{(n)})\,.
\end{eqnarray} 
Taking the variation of the action $\delta S=0$ leads to the Euler-lagrange 
equation of motion
\begin{eqnarray}
\sum_{i=0}^n (-1)^i\frac{d^i}{dt^i}\Bigl(\frac{\partial {\cal L}}{\partial 
q^{(i)}}\Bigr)=0\,.
\end{eqnarray}
The hamiltonian is obtained by introducing $n$ generalized momenta
\begin{eqnarray}
p_i=\sum_{j=i+1}^n (-1)^{j-i-1}\frac{d^{j-i-1}}{dt^{j-i-1}}
\Bigl(\frac{\partial {\cal L}}{\partial q^{(j)}}\Bigr)\,,
~~i=1,\cdots,n,
\end{eqnarray}
or
\begin{eqnarray}
p_n=\frac{\partial {\cal L}}{\partial q^{(n)}}\,,~~
p_i=\frac{\partial {\cal L}}{\partial q^{(i)}}-\frac{d}{dt}p_{i+1}\,,~~i=1,
\cdots,n-1,
\label{canonical momenta}
\end{eqnarray}
and $n$ independent variables
\begin{eqnarray}
q_1\equiv q\,,~~
q_i\equiv q^{(i-1)}\,,~~i=2,\cdots,n\,.
\end{eqnarray}
The lagrangian now depends on the $n$ coordinates $q_i$ and on the first 
derivative $\dot{q}_n=q^{(n)}$.
The hamiltonian is defined as
\begin{eqnarray}
{\cal H}(q_i,p_i)=\sum^n_{i=1}p_i\dot{q}_i-{\cal L}=\sum^{n-1}_{i=1} p_i 
q_{i+1}+p_n \dot{q}_n-{\cal L}\,.
\end{eqnarray}
The canonical equations of motion turn out to be
\begin{eqnarray}
\dot{q}_i=\frac{\partial {\cal H}}{\partial p_i}\,,~~\dot{p}_i=-\frac{\partial {\cal H}}{\partial q_i}\,.
\end{eqnarray}
Thus, we replace a theory of one coordinate $q$ system obeying $2n-$th 
differential equation
with a set of 1-st order canonical equations for $2n$ phase-space 
variables $[q_i,p_i]$.

We consider a simple example including a second derivative term \cite{simon}, 
defined as 
\begin{eqnarray}
{\cal L}=\frac{1}{2}(1+\varepsilon^2\omega^2)\dot{q}^2-\frac{1}{2}
\omega^2q^2-\frac{1}{2}\varepsilon^2\ddot{q}^2\,,
\end{eqnarray}
where constant $\epsilon$ works as a coupling constant of second 
derivative term. 
The equation of motion is
\begin{eqnarray}
(1+\varepsilon^2 \omega^2)\ddot{q}+\omega^2 q+\varepsilon^2 
q^{(4)}=0\,.
\label{eq_quanta}
\end{eqnarray}
From Eq.~(\ref{canonical momenta}), one gets 
\begin{eqnarray}
\pi_{\dot{q}}=\frac{\partial{\cal L}}{\partial \ddot{q}}
=-\varepsilon^2\ddot{q}\,,~~
\pi_q=\frac{\partial{\cal L}}{\partial \dot{q}}-\frac{d}{dt}
\Bigl(\frac{\partial{\cal L}}{\partial \ddot{q}}\Bigr)
=(1+\varepsilon^2\omega^2)\dot{q}+\varepsilon^2\dddot{q}\,.
\end{eqnarray}
Thus the hamiltonian becomes 
\begin{eqnarray}
{\cal H}&=&\dot{x}\pi_q+\ddot{q}\pi_{\dot{q}}-{\cal L} 
\nonumber \\
&=&\dot{q}\pi_q-\frac{1}{2\varepsilon^2}\pi_{\dot{q}}^2
-\frac{1}{2}(1+\varepsilon^2\omega^2)\dot{q}^2+\frac{1}{2}\omega^2q^2\,.
\end{eqnarray}
We introduce the new canonical variables
\begin{eqnarray}
&&q_+=\frac{1}{\omega\sqrt{1-\varepsilon^2\omega^2}}
(\varepsilon^2\omega^2\dot{q}-\pi_q),~~
p_+=\frac{w}{\sqrt{1-\varepsilon^2\omega^2}}(q-\pi_{\dot{q}})\,, \nonumber \\
&&q_-=\frac{\varepsilon}{\sqrt{1-\varepsilon^2\omega^2}}(\dot{q}-\pi_q),~~
p_-=\frac{1}{\varepsilon\sqrt{1-\varepsilon^2\omega^2}}(\varepsilon^2\omega^2 
q-\pi_{\dot{q}})\,,
\nonumber 
\end{eqnarray}
and the hamiltonian can be written in terms of these variables as 
\begin{eqnarray}
{\cal H}\to \frac{1}{2}(p_+^2+\omega^2 q_+^2)-\frac{1}{2}(p_-^2
+\frac{1}{\varepsilon^2} q_-^2)\,. \nonumber 
\end{eqnarray}The corresponding energy spectra is then given by
\begin{eqnarray}
E=(n+\frac{1}{2})\omega-(m+\frac{1}{2})\frac{1}{\varepsilon}~,~~n,m=0,1,2,\cdots
\end{eqnarray}
One can see that in the limit $\epsilon\to 0$ the energy goes to 
negative infinity rather than approaching to the harmonic oscillator 
energy eigenstates.

To obtain physically meaningful solutions, we employ the perturbative 
analysis where the solution is expanded in terms of the small coupling 
constant and the Euler-Lagrange equation of motion is replaced with the 
corresponding perturbative equation. 
The solutions of the equations of motion that are ill behaved in the limit 
$\epsilon\to 0$ are excluded from the very 
beginning~\cite{eliezer89,jaen,simon}.

We assume that the solution of Eq.(\ref{eq_quanta}) can be written as
\begin{eqnarray}
q_{\rm pert}(t)=\sum^{\infty}_{n=0} \epsilon^n q(t)\,. \label{q}
\end{eqnarray}
Substituting Eq.(\ref{q}) into Eq.(\ref{eq_quanta}) and taking time 
derivatives of these equations, we obtain the constraints for higher 
derivative terms
\begin{eqnarray}
&&O(\epsilon^0)  \nonumber \\
&&\hspace{3mm}equation :~~\ddot{q}_0+\omega^2q_0=0\,, \\
&&\hspace{3mm}constraints :~~\dddot{q}_0=-\omega^2\dot{q}_0, \ddddot{q}_0=\omega^4 q_0\,. \label{cnst0} \\
&&O(\epsilon^2)  \nonumber \\
&&\hspace{3mm}equation :~~\ddot{q}_2+\omega^2\ddot{q}_0+\omega^2 q_2
+\ddddot{q}_0=0\,,  \nonumber \\
&&\hspace{13mm}\Rightarrow \ddot{q}_2+\omega^2q_2=0\,,~~~~
({\rm using}~(\ref{cnst0}))\,, \\
&&\hspace{3mm}constraints :~~\dddot{q}_2=-\omega^2\dot{q}_2, 
\ddddot{q}_2=\omega^4 q_2\,. \label{cnst2} \\
&&O(\epsilon^4)  \nonumber \\
&&\hspace{3mm}equation :~~\ddot{q}_4+\omega^2\ddot{q}_2+\omega^2 q_4
+\ddddot{q}_2=0\,,  \nonumber \\
&&\hspace{13mm}\Rightarrow \ddot{q}_4+\omega^2q_4=0\,,
~~~~({\rm using}~(\ref{cnst2}))\,, \\
&&\hspace{3mm}constraints :~~\dddot{q}_4=-\omega^2\dot{q}_4, 
\ddddot{q}_4=\omega^4 q_4\,. \label{cnst4}
\end{eqnarray}
Combining these results, we find the perturbative equation of 
motion up to $O(\epsilon^4)$
\begin{eqnarray}
\ddot{q}_{\rm pert}+\omega^2q_{\rm pert}=O(\epsilon^6)\,.
\end{eqnarray}
which is the equation for harmonic oscillator. 

\section{\label{sec:level6}Search for the soliton solutions (2) -- 
perturbative analysis --\protect\\}

Let us employ the perturbative method introduced in the last section to our 
problem. We assume that $g_2$ is relatively small and can be considered as 
a perturbative coupling constant. Thus, the perturbative solution is written 
by a power series in $g_2$
\begin{eqnarray}
w(\eta)=\sum^\infty_{n=0}g_2^n w_n(\eta)\,.
\label{sol_expand}
\end{eqnarray}
Substituting Eq.(\ref{sol_expand}) into Eq.(\ref{fsn_eq2}), we obtain 
the classical field equation in $O(g^0_2)$
\begin{eqnarray}
f_0(w_0,w_0',w_0'')+g_1f_1(w_0,w_0',w_0'')=0\,. \label{classical}
\end{eqnarray}
Taking derivatives for both sides in Eq.(\ref{classical}) and solving 
for $w_0'',w_0^{(3)},w_0^{(4)}$ read the following form
of the constraint equations for higher derivatives
\begin{eqnarray}
w_0^{(i)}=F^{(i)}(w_0,w_0')\,,~~i=2,3,4\,.
\label{constraint}
\end{eqnarray}
The equation in $O(g^1_2)$ can be written as 
\begin{eqnarray}
(f_0+g_2 f_1)_{O(g^1_2)}+f_2(w_0,w_0',w_0'',w_0^{(3)},w_0^{(4)})=0\,.
\label{fsn_eq21}
\end{eqnarray}
Substituting the constraint equations (\ref{constraint}) into 
Eq.(\ref{fsn_eq21})
and eliminate the higher derivative terms, one can obtain the perturbative 
equation of motion
\begin{eqnarray}
f_0(w,w',w'')+g_1 f_1(w,w',w'')+g_2 \tilde{f}_2(w,w')=O(g^2_2)\,.
\label{fsn_eq2_p}
\end{eqnarray}
One can see that Eq.(\ref{fsn_eq2_p}) has stable soliton solutions. 

\begin{figure}
\includegraphics[height=9cm, width=12cm]{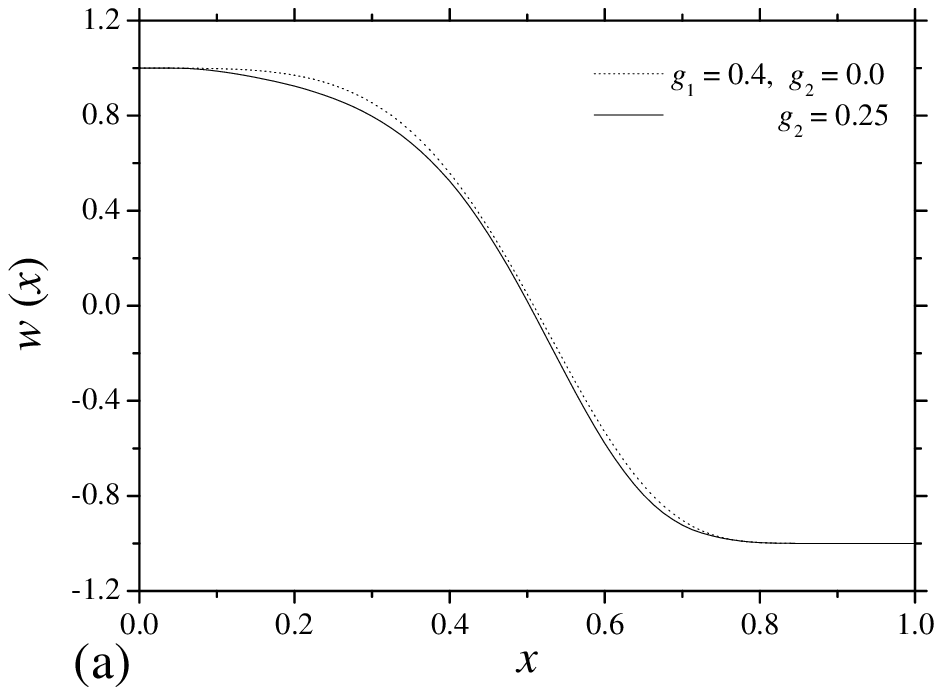}
\includegraphics[height=9cm, width=12cm]{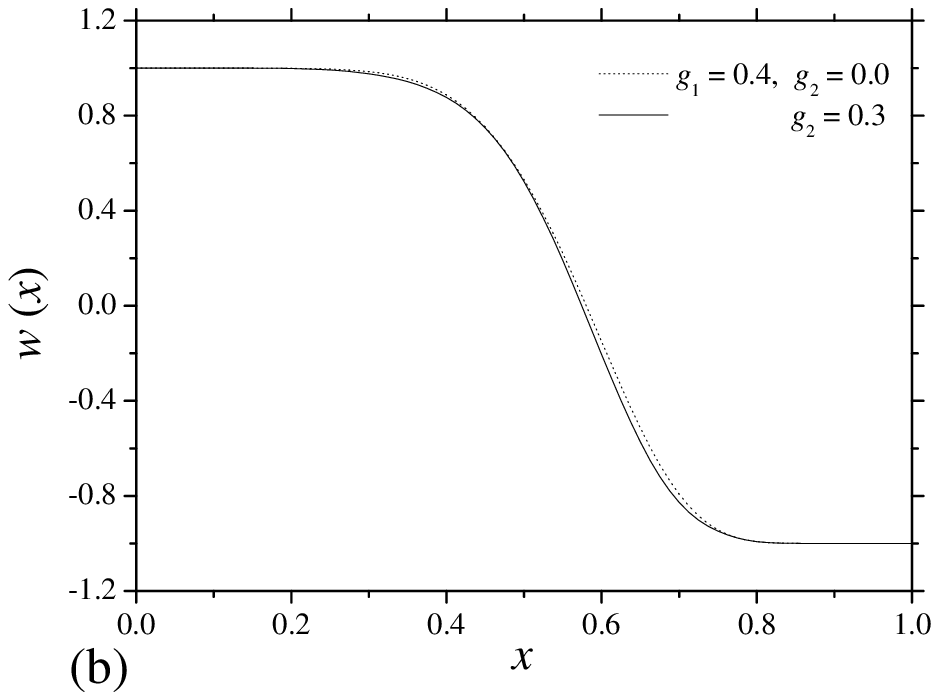}
\caption{\label{fig:Fig2} The function $w(x)$ of (a) $H=1,g_1=0.4,g_2=0.25$, (b) $H=2,g_1=0.4,g_2=0.3$
(the rescaling radial coordinate $x=\eta/(1-\eta)$ is used), together 
with the results of the original SFN model. }
\end{figure}

\begin{figure}
\includegraphics[height=9cm, width=13cm]{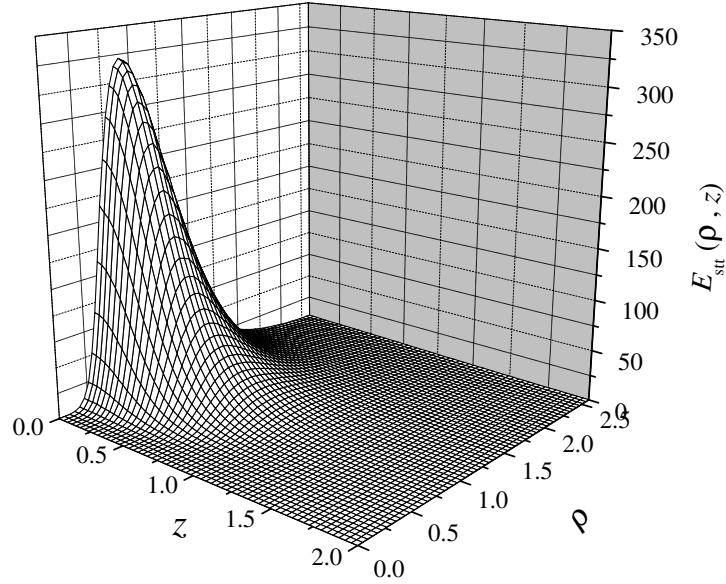}
\caption{\label{fig:Fig3}The energy density of naive SFN (\ref{fsn_ac}) model
in the cylindrical coordinates $(\rho,z)$ for $H=2,g_1=0.4$.}
\end{figure}

\begin{figure}
\includegraphics[height=9cm, width=13cm]{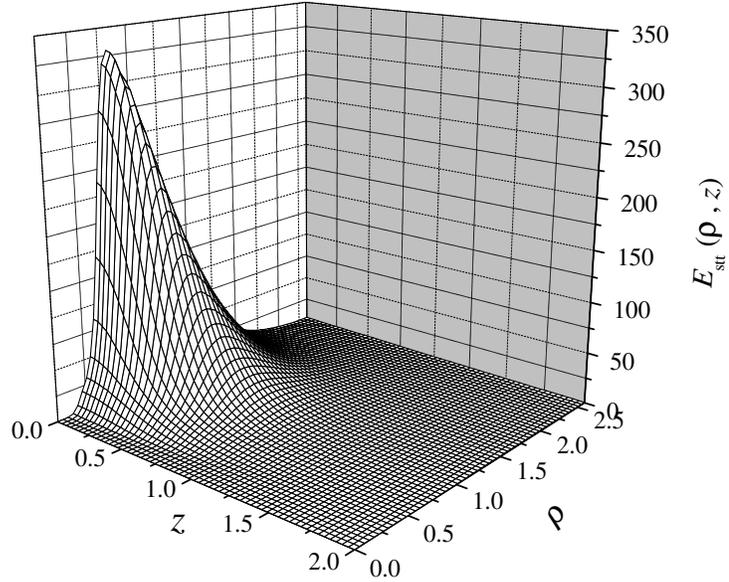}
\caption{\label{fig:Fig4}The energy density of extended action (\ref{fsn_ac2})
in the cylindrical coordinates $(\rho,z)$ for $H=2,g_1=0.4,g_2=0.3$.}
\end{figure}

\begin{figure}
\includegraphics[height=9cm, width=12cm]{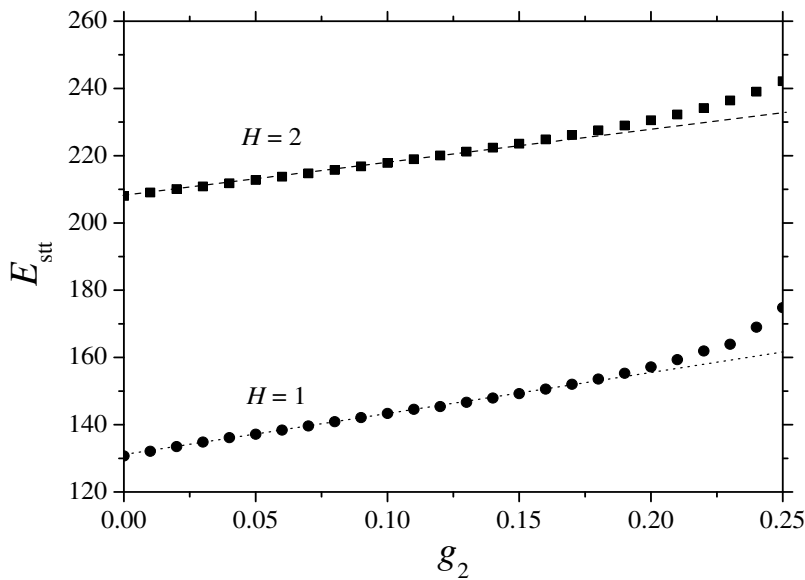}
\caption{\label{fig:Fig5} The static energies of the soliton for $H=1,2$ 
as a function of $g_2$ ($g_1=0.4$).
Also the linearly fitted line of the $H=1$ : $E_{\rm stt}=131+121g_2$  
(dotted line) and 
of the $H=2$ : $E_{\rm stt}=208+102g_2$  (dashed line).}
\end{figure}

\begin{figure}
\includegraphics[height=9cm, width=12cm]{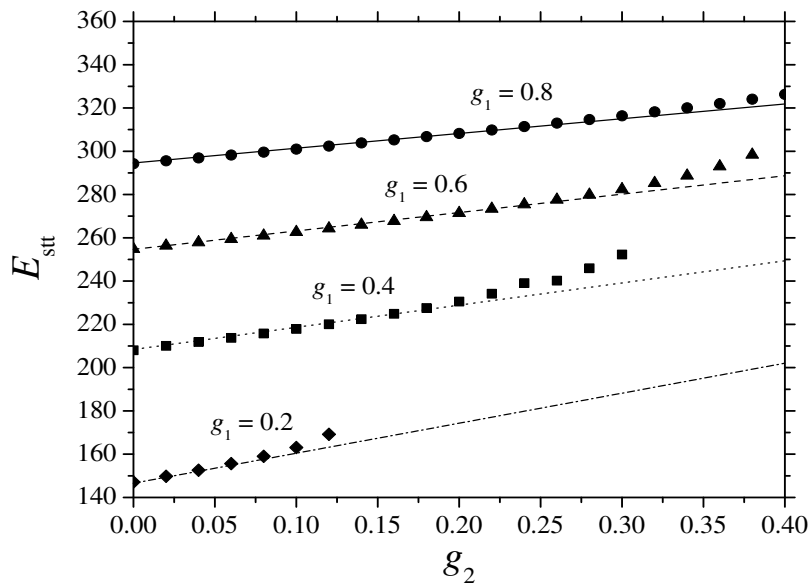}
\caption{\label{fig:Fig6} The static energies of the soliton for 
$g_1=0.2,0.4,0.6,0.8$ 
as a function of $g_2$ ($H=2$). 
Also the linearly fitted line of the $g_1=0.2$ : 
$E_{\rm stt}=147+142g_2$  (dot-dashed line), 
the $g_1=0.4$ : $E_{\rm stt}=208+102g_2$  (dotted line), 
the $g_1=0.6$ : $E_{\rm stt}=255+83g_2$  (dashed line) and 
of the $g_1=0.8$ : $E_{\rm stt}=294+72g_2$  (solid line).}
\end{figure}

\begin{figure}
\includegraphics[height=9cm, width=12cm]{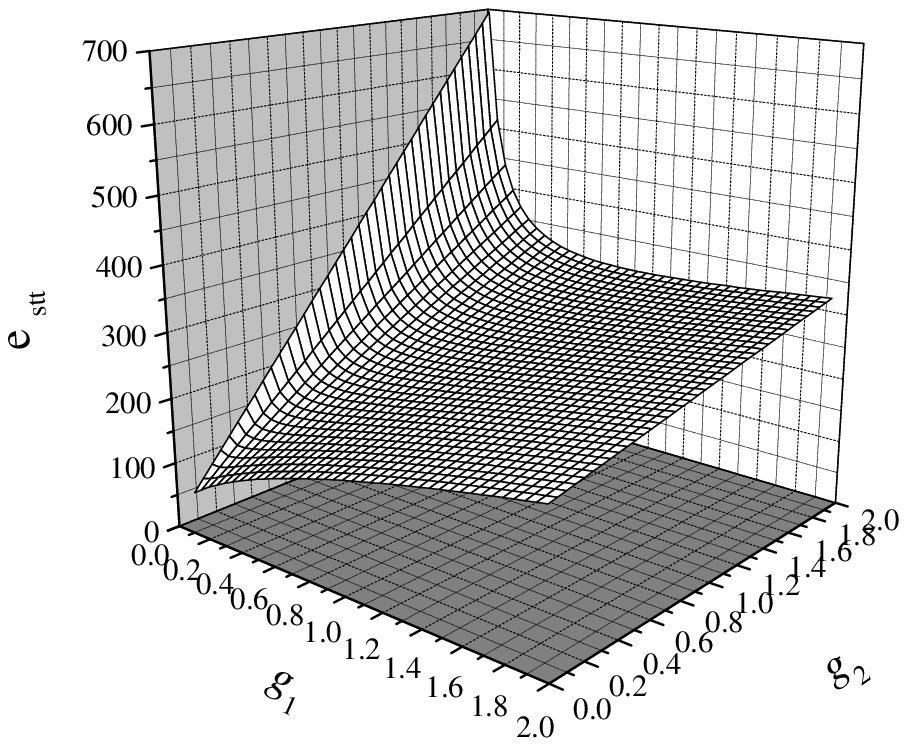}
\caption{\label{fig:Fig7} The plot of the conjecture (\ref{lowerbound2}) 
for $(g_1,g_2)$, in the case of $H=1$.}
\end{figure}

The numerical results of the functions $w(\eta)$ are displayed in Fig.\ref{fig:Fig2}. 
In Fig.\ref{fig:Fig3}, we show the energy density plot for the original SFN model 
in the cylindrical coordinates $(\rho,z)$. In Fig.\ref{fig:Fig4}, we present the 
energy density for our extended soliton in the $(\rho,z)$. Both results 
share the toroidal shape and no notable 
difference at least for the small coupling constant $g_2$.
In Fig.\ref{fig:Fig5}, we plot the total energy for $H=1,2,g_1=0.4$ as a 
function of the coupling constant $g_2$.
As can be seen, the change is moderate and, more interestingly, the energy  
seems to be linearly dependent on $g_2$ at the region of smaller $g_2$. 
For larger $g_2$, the data gradually 
deviate from the linear behavior. The deviation would be due to the 
fact that our analysis is based on the first order perturbation.  
Therefore, it is possible to observe the critical values of $g_2$ for 
each $g_1$ and $H$ in which our simple first order perturbation 
is valid. Performing linear fitting, one finds 
\begin{eqnarray}
&&E_{H=1}\sim 131+121g_2\,, \nonumber \\
&&E_{H=2}\sim 208+102g_2\,. \nonumber 
\end{eqnarray}
In Fig.\ref{fig:Fig6}, we plot the energies for various values 
of $g_1$ as a function of $g_2$.
They are fitted to the following linear functions 
\begin{eqnarray}
&&E_{g_1=0.2}\sim 147+142g_2\,,\nonumber \\
&&E_{g_1=0.6}\sim 255+83g_2\,,\nonumber \\
&&E_{g_1=0.8}\sim 294+72g_2\,.\nonumber
\end{eqnarray}
From these results we can extract formulation of topological bound for the energy 
of the second derivative term $E_{\rm 2nd}$:
\begin{eqnarray}
E_{\rm 2nd}=\beta \frac{g_2}{\sqrt{g_1}}H^{-1/4}\,.
\end{eqnarray}
If $\beta\sim 76$ is chosen, all data are well fitted by this single 
formula. For the energy from standard SFN action, we tentatively 
employ the topological lower bound~(\ref{lowerbound}). Then the topological 
lower bound including second derivative terms becomes 
\begin{eqnarray}
E_{\rm stt}\ge {\rm e}_{\rm stt}=\alpha \sqrt{g_1}H^{3/4}+\beta
\frac{g_2}{\sqrt{g_1}} H^{-1/4}
\label{lowerbound2}
\end{eqnarray}
with $\alpha=16\pi^2$. Though this formula is not based on any theory,  
it contains interesting information about the spectra of soliton.  
That is , (\ref{lowerbound2}) has the minima at
\begin{eqnarray}
\frac{\partial {\rm e}_{\rm stt}}{\partial g_1}=0~\to~
g_1=\frac{\beta}{\alpha}\frac{g_2}{H}
\label{extermum2}
\end{eqnarray}
for given $g_2$. 
The energy bound is thus ${\rm e}_{\rm stt}=2\alpha\sqrt{g_1} H^{3/4}$. 

Let us summarize the numerical results presented in this section. 
If we take into account the second derivative 
term, the obtained soliton mass will become twice value of the one in 
the naive SFN soliton. In Fig.\ref{fig:Fig7}, a schematic plot of 
Eq.(\ref{lowerbound2}) is shown, and the existence of such local minima is 
observed. Interestingly, though at first glance our lower bound formula 
(\ref{lowerbound2}) has no 3/4-scaling behavior like (\ref{lowerbound}) 
essential to knotted solitons 
\cite{lin04}, the energy at local minima recovers this scaling law.

Unfortunately, such minima can not be observed at present within 
our numerical framework. Because, from the condition (\ref{extermum2}), 
the value of $g_2$ should be almost twice of $g_1$ for $H=1$, 
and four times for $H=2$. Otherwise, we need to take into account the higher order 
perturbation terms to employ totally different formalism  to achieve these values. 

\section{\label{sec:level7}Summary\protect\\}

In this paper we have studied the Skyrme-Faddeev-Niemi
model and its extensions by introducing the reduction scheme of the SU(2) 
Yang-Mills theory to the corresponding low-energy effective model.
To simplify the matter, we proposed an ansatz for $\bm{n}$. That extensively 
reduced the computational time and did not affect to the property of the soliton 
solution.
The requirement of consistency between the low-energy effective actions 
of the YM and the SFN type model leads us to take into account second 
derivative terms in the action. 
However, we found that such an action including the second derivative terms 
does not have stable soliton solutions.
This is due to the absence of the energy bound in higher 
derivative theories. 
This fact inspired us to employ the perturbative analysis to our effective 
action. Within the perturbative analysis, we were able to obtain soliton 
solutions. 

It should be noted that our solutions do not much differ from 
the solution of original SFN model, at least in the perturbative regime. 
We suspect that an appropriate truncation (for instance ``extra fourth order term 
+ second derivative term'') can supply stable solutions that are 
close to the original SFN model. Thus we conclude that the topological 
soliton model consisting of the ``kinetic term + a special fourth order 
term'' like SFN model is a good approximation. 

Our analysis is based on perturbation and the coupling constant $g_2$ is 
assumed to be small. 
However, Wilsonian renormalization analysis of YM theory~\cite{gies01} 
suggests that the coupling constants $g_1,g_2$ (and the mass scale parameter 
$\Lambda$) depend on the renormalization group time $t=\log k/\Lambda$ 
($k,\Lambda$ are infrared, ultraviolet cutoff parameter) and they are 
almost comparable. To improve the analysis, we could perform the 
next order of perturbation, but it is tedious and spoils the simplicity 
of the SFN model unfortunately. 

We found numerically that the energy of the extended soliton solutions 
is linear in the coupling constant $g_2$ and then extracted a new 
mass formula for the soliton solutions including second derivative term.
We expect that the global minima for the coupling constant $g_2$ exists 
and the corresponding energy becomes twice the value of the naive SFN action. 
Of course, this statement is not based on any theory and 
we have to wait for its theoretical confirmation.  
Since our mass formula was obtained from numerical study in the 
perturbative approach, it is uncertain whether such linear behavior 
is kept for larger coupling constant $g_2$
({\it like} twice of $g_1$ for $H=1$ and four times for $H=2$). To 
confirm that, we should proceed to investigate next order perturbation, 
or, otherwise, find some analytical evidence of this formula. 
We point out that the perturbative treatment is only used for 
excluding the ill behavior of the second derivative field theory. 
We believe that applying this prescription should not alter the 
essential feature of the soliton solutions, {\it e.g.}, existence 
of the solutions, linear behavior of the mass spectra, {\it et.al.}.

Finally, let us mention the application of the soliton solutions 
to the glueball. Obviously this is one of the main interest to study 
the model and, many authors have given discussions about 
this \cite{gies01,faddeev04,cho04}. On the other hand, 
the possibility of the magnetic condensation of the QCD vacuum within 
the Cho-Faddeev-Niemi-Shabanov decomposed Yang-Mills theory have been 
studied by Kondo\cite{kondo}. 
The author claims the existence of the nonzero off diagonal 
gluon mass $M_X$, which is induced in terms of the 
condensation of the magnetic potential part of the decomposition 
$\bm{B}_\mu\sim (\partial_\mu\bm{n})\times\bm{n}$, as 
\begin{eqnarray}
M_X^2=\langle \bm{B}_\mu\cdot\bm{B}_\mu\rangle=\langle
 (\partial_\mu\bm{n})^2\rangle\,.
 \end{eqnarray} 
Throughout our calculation, we set the overall coupling constant 
$\Lambda=1$ in the action (\ref{fsn_ac}) but, in this sense, 
it should reflect the information of such gluon mass, or the 
condensation property of the vacuum. After a careful examination of 
the value of the property of $\Lambda$, we will be able to accomplish 
the detailed predictions for the glueball mass. 
 
\section*{Acknowledgments}
 
We are grateful to Kei-Ichi Kondo for drawing our attention to 
this subject and for many useful advices. We also thank M.Hirayama and 
J.Yamashita for valuable discussions.

\end{document}